\documentclass[aps,prd,nofootinbib,amsmath,amssymb]{revtex4}
\usepackage{eurosym}
\usepackage{amsmath,amssymb}
\usepackage{graphicx}
\usepackage{dcolumn}
\usepackage{bm}
\usepackage{color}
\usepackage{multirow}

\usepackage{array}
\usepackage{booktabs}
\usepackage{placeins}
\usepackage[detect-all]{siunitx}
\usepackage{makecell}
\usepackage{xcolor}
\usepackage{yfonts}
\graphicspath{ {/Users/Nastia/Desktop/research_dis/papers_writing/xenes/nov_5} }

\newcolumntype{P}[1]{>{\centering\arraybackslash}p{#1}}
\begin{document}
\title{$\Omega NN$ and $\Omega\Omega N$ systems

with the updated HAL QCD $\Omega N$ Interaction}

\author{R. Lazauskas$^{1}$, R. Ya. Kezerashvili$^{2,3}$, I. Filikhin$^{4}$}
\affiliation{$^{1}$IPHC, CNRS/IN2P3, Universit\'e de Strasbourg, 67037 Strasbourg, France\\
$^{2}$New York City College of Technology, The City University of New York,
Brooklyn, NY, USA\\
$^{3}$The Graduate School and University Center, The City University of New
York, New York, NY, USA\\
$^{4}$North Carolina Central University, Durham, NC, USA 
}

\begin{abstract}

Using the updated physical-point lattice QCD $\Omega N$ interaction from the HAL QCD Collaboration, which supports a shallow state near unitarity, within configuration-space Faddeev equations, we calculate the binding energies and spatial properties of three-body $\Omega NN$ ($T=0, 1$) and $\Omega\Omega N$ ($J^\pi=1/2^+, 5/2^+$) systems, including Coulomb interactions for charged configurations. Bound states emerge across all investigated channels with distinct binding energies and spatial distributions. Compared to earlier HAL QCD potentials, the updated interaction yields noticeably weaker binding, whereas Coulomb attraction enhances binding in charged states. These results provide benchmark theoretical predictions for strangeness $S=-3$ and $-6$ three-baryon systems and highlight their sensitivity to physical-point lattice QCD $\Omega N$ interaction.

\end{abstract}

\date{\today }
\maketitle

\textbf{Introduction.}
The existence of a bound $N\Omega$ dibaryon was first suggested
nearly four decades ago in the framework of constituent-quark
models~\cite{Goldman1987}. Since then, dibaryons containing $\Omega$ baryons have been studied extensively, particularly the $N\Omega$ in the
${}^{5}S_2$ channel and $\Omega\Omega$ in the ${}^{1}S_0$ channel. These systems are considered among the most promising candidates for strange dibaryon states (see the reviews~\cite{Clement2017,Cho2017}). Subsequent studies have examined the effects of color-dependent interactions, hidden-color configurations, quark delocalization, color screening, and coupled-channel dynamics~\cite{Oka1981,Harvey1981a,Pang2004,Li2001,
SekiharaHashiguchi2023}. Chiral effective field theory~\cite{Haidenbauer2020,Haidenbauer2017} and a
phenomenological meson-exchange model~\cite{Sekihara2018} provide complementary descriptions of the $\Omega N$ interaction.

Lattice quantum chromodynamics (QCD) and relativistic heavy-ion collisions offer complementary ways to investigate baryon interactions. QCD provides a first-principles framework for studying these interactions in terms of the  quark and gluon dynamics. In particular, the HAL QCD method makes it possible to extract effective baryon–baryon potentials from lattice-QCD correlation functions~\cite{Aoki2012}. This method has been applied successfully to a range of hadronic systems, including systems involving decuplet baryons such as $\Omega N$ ~\cite{Etminan2014,Iritani2019,Zhang2026QCD2026} and $\Omega\Omega$~\cite{OmegaOmega2018}. The resulting potentials provide valuable input for studies of bound states and resonances in few-baryon systems.

Experimental constraints on the $\Omega N$ interaction have also emerged from femtoscopic measurements of proton-$\Omega$ momentum correlations in relativistic heavy-ion collisions. The sensitivity of these correlations to the strong interaction between hadrons has been demonstrated theoretically in Refs.~\cite{Morita2016,Morita2020}. Measurements by the START Collaboration at the RHIC, BNL, of the $p \Omega^-$-correlation function in Au$+$Au collisions at %√sN N = 200
~\cite{STAR2019} and proton-proton collisions at the CERN LHC ~\cite{ALICE2020} (ALICE Collaborations) provides a precise method  provide evidence consistent with a substantial attractive interaction, stimulating further theoretical investigations of $N\Omega$ correlations and dibaryon production~\cite{Zhang2020,Yan2025}. Nevertheless, the microscopic origin of this attraction and the roles played by different interaction mechanisms remain incompletely understood. Establishing the connection between quark-level interactions, effective hadronic descriptions, and lattice-QCD-derived potentials is therefore important for understanding the nature of the predicted near-threshold $N\Omega$ state.

The $\Omega\Omega$ dibaryon has been investigated
within various quark models, which predict either
a strongly attractive~\cite{Zhang2000,Li2001,Dai2006}
or weakly repulsive~\cite{Wang1995,Buchoff2012}
$\Omega\Omega$ interaction.
Its possible production in relativistic heavy-ion
collisions has also been explored using a multi-phase transport model~\cite{Yu2001,Yu2002,Pal2005}.
More recently, HAL QCD calculations near the
physical pion mass ($m_\pi=146$ MeV) predicted
a shallow $\Omega\Omega$ bound state with binding
energies of approximately $1.6$ and $0.7$ MeV
without and with Coulomb repulsion,
respectively~\cite{OmegaOmega2018}.
Together, these studies provide a framework for investigating exotic dibaryons and multibaryon systems with nonzero strangeness.

The HAL QCD Collaboration recently reported an updated
$\Omega N$ interaction in the ${}^{5}S_2$ channel,
obtained directly at the physical pion mass
$m_\pi \simeq 137$ MeV~\cite{Zhang2026QCD2026},
improving upon the earlier near-physical-point calculation
at $m_\pi \simeq 146$ MeV~\cite{Iritani2019}. Both studies by the HAL QCD Collaboration investigate
the $N\Omega$ dibaryon in the $S$-wave, spin-2
(${}^{5}S_2$) channel using $(2+1)$-flavor lattice QCD
and the time-dependent HAL QCD method.
The recent study by Zhang \textit{et al.}~\cite{Zhang2026QCD2026}
represents a substantial improvement over the earlier
calculation by Iritani \textit{et al.}~\cite{Iritani2019}.
The principal change is the transition from a
near-physical quark-mass simulation (K-computer ensemble,
$m_\pi \simeq 146$ MeV) to a physical-point simulation
(Fugaku ensemble, $m_\pi \simeq 137.1$ MeV),
with substantially increased statistics.
Additional methodological improvements include the
Misner $S$-wave projection, direct determination of the
bound-state pole through the Lippmann--Schwinger equation,
and a finite-charge-size treatment of the Coulomb interaction.
The updated calculation predicts a shallower $N\Omega$
bound state, with a binding energy of $0.61$ MeV at $t/a=14$,
compared with approximately $1.54$ MeV in the earlier study \cite{Iritani2019}.
Including Coulomb attraction gives a $p\Omega^-$ binding
energy of $1.25$ MeV, compared with approximately
$2.46$ MeV in the earlier calculation.
Thus, the central values of the binding energies are reduced by approximately a factor of two, although the predictions remain compatible within the quoted uncertainties. Since the $N\Omega({}^{5}S_2)$ system lies close to the unitary limit, these differences motivate an investigation of how the updated interaction affects few-body binding energies.

The binding energy of the lightest $\Omega NN$ and $\Omega \Omega N$ systems have been investigated using Faddeev equations in momentum ~\cite{GV2016,GV0,GV} and configuration~\cite{FilKezPRC2025,FilKezPRD2025,FilKezPRC2026} space, with HAL QCD $\Omega N$ \cite{Iritani2019} and $\Omega \Omega$ \cite{OmegaOmega2018} interactions and with meson-exchange potential~\cite{Sekihara2018}. The same interactions have subsequently been employed in Refs.~\cite{Zhang2022,ESE2023} to calculate the binding energy of the $\Omega NN$ system using different theoretical approaches. In Ref. \cite{Etminan2026} the $\Omega^- pn$ system is considered in the isospin-spin-parity channel $I(J^P)=0(5/2^+)$ using HAL QCD $\Omega N$ \cite{Zhang2026QCD2026} and Malfliet–Tjon \cite{Malfliet1969,MTcorr} $pn$ potentials using the Gaussian Expansion Method \cite{Hiyama2003GEM}. %GEM is a variational approach in which the total wave function is expanded in terms of Gaussian basis functions associated with the relevant Jacobi coordinate sets. 
All these studies suggest that the $\Omega NN$ system may be more strongly bound than the $\Omega N$ subsystem. Such an enhancement would reflect the combined effects of the attractive $\Omega N$ interaction and the additional binding generated by the three-body dynamics.

In the present work, we investigate the three-body   $\Omega NN$ and
$\Omega\Omega N$  for the first time using the updated physical-point HAL QCD $\Omega N$ interactio \cite{Zhang2026QCD2026} within the configuration-space Faddeev
framework. 
We compare the binding energies and spatial properties obtained with this interaction with the corresponding results obtained using the earlier near-physical-point HAL QCD $\Omega N$~\cite{Iritani2019}. The HAL QCD $\Omega\Omega$ interaction 
~\cite{OmegaOmega2018} is used in the study of the $\Omega\Omega N$ system. We examine the effects of Coulomb attraction and of adding a third baryon on the binding and structural properties of these systems. In particular, we investigate whether the deeply bound three-body states remain bound despite the reduced two-body attraction, and how the weakly bound states respond to changes in the interaction and approach the corresponding breakup thresholds.

\iffalse
In the present study, we investigate the $\Omega NN$ and
$\Omega\Omega N$ systems using the recently updated
HAL QCD $\Omega N$ interaction, together with the earlier
$\Omega N$ potentials from
Refs.~\cite{Sekihara2018,Iritani2019},
the HAL QCD $\Omega\Omega$ interaction, and the
Malfliet--Tjon MT I--III $NN$
potential~\cite{Malfliet1969,MTcorr}.
By comparing lattice-QCD-derived and meson-exchange
$\Omega N$ interactions, we assess the sensitivity of
few-body binding energies and spatial structures to
the choice of interaction model.
We also examine the effects of Coulomb attraction
and the addition of third baryon on the binding and
structural properties of these systems.
\fi

\begin{table*}[!b]
\centering
\caption{
Parameters of the HAL QCD $\Omega N$ \cite{Zhang2026QCD2026} and $\Omega\Omega$ \cite{OmegaOmega2018} 
interaction potentials.
 part of the table lists the parameters of the 
$\Omega N$ interaction in the ${}^{5}S_2$ channel
at Euclidean time slices $t/a=12$--$16$ together with corresponding binding energies of
$\Omega n$ and $\Omega p$ states, calculated using central values of the parameter sets.
The lower section gives the parameters of the
$\Omega\Omega$ interaction in the ${}^{1}S_0$ channel
at $t/a=17$ and corresponding binding energy of di-$\Omega$ state.
Statistical uncertainties are given in parentheses.
}
\label{tab:Omega_potentials2}
%\begin{ruledtabular}
\begin{tabular}{cccccccc}
\hline\hline
\multicolumn{8}{c}{
$\Omega N$ interaction (${}^{5}S_2$)
}\\
\hline
$t/a$
& $a_1$ [MeV]
& $b_1$ [fm$^{-2}$]
& $a_2$ [MeV]
& $b_2$ [fm$^{-2}$]
& $a_3$ [MeV$\cdot$fm$^2$] 
& $B_{\Omega n}$ [MeV] 
& $B_{\Omega p}$ [ MeV] \\ 
\hline
12 & $-312.7(2.2)$  & $51.8(1.1)$
   & $-76.3(7.5)$   & $1.44(4)$
   & $-289.0(46.9)$ & $1.2320$  & $1.9572$ \\
13 & $-314.8(3.9)$  & $50.6(2.3)$
   & $-83.9(11.3)$  & $1.43(8)$
   & $-244.1(78.1)$ & $1.0194$  & $1.7219$ \\
14 & $-311.5(6.7)$  & $49.0(2.4)$
   & $-94.7(16.6)$  & $1.38(10)$
   & $-162.6(111.1)$ & $0.6179$  & $1.2571$\\
15 & $-307.1(15.6)$ & $52.6(4.4)$
   & $-76.2(17.2)$  & $1.47(19)$
   & $-265.3(118.3)$ & $0.6840$  & $1.3240$ \\
16 & $-285.4(24.1)$ & $56.4(8.9)$
   & $-56.3(17.3)$  & $1.73(46)$
   & $-358.8(131.5)$  & $0.4234$  & $0.9875$\\

\hline
\multicolumn{8}{c}{
$\Omega\Omega$ interaction (${}^{1}S_0$),
$t/a=17$
}\\
\hline

$c_1$ [MeV]
& $d_1$ [fm]
& $c_2$ [MeV]
& $d_2$ [fm]
& $c_3$ [MeV]
& $d_3$ [fm] 
& $B_{\Omega \Omega}$[ MeV]
&\\
\hline

$914(52)$
& $0.143(5)$
& $305(44)$
& $0.305(29)$
& $-112(13)$
& $0.949(58)$ 
& 0.5977  &\\
\hline\hline
\end{tabular}
%\end{ruledtabular}
\end{table*}
\textbf{Interaction.}
A theoretical description of the $\Omega NN$ and
$\Omega\Omega N$ systems requires the $\Omega N$, $NN$,
and $\Omega\Omega$ interactions. In the present calculations, we use the HAL QCD $\Omega N$
interactions~\cite{Iritani2019,Zhang2026QCD2026}
together with the  $\Omega\Omega$ interaction~\cite{OmegaOmega2018}.
These potentials were extracted using the HAL QCD
method from lattice-QCD calculations.

The recently updated HAL QCD $\Omega N$ interaction
is represented by the fitted potential~\cite{Zhang2026QCD2026}
\begin{equation}
V_{\mathrm{fit}}(r)
=
a_1 e^{-b_1 r^2}
+
a_2 e^{-b_2 r^2}
+
a_3
\left(
1-e^{-(r/R)^2}
\right)
\left(
\frac{e^{-m_\pi r}}{r}
\right)^2,
\label{tab:Omega_potentials}
\end{equation}
where the parameters $a_1$, $a_2$, $a_3$, $b_1$, $b_2$ ,$b_3$
in the ${}^{5}S_2$ channel at Euclidean time slices
$t/a=12$--$16$ are listed in
Table~\ref{tab:Omega_potentials2}.
The physical pion mass is $m_\pi=137.1$ MeV,
and $R=2.0$ fm.

For the $\Omega\Omega$ interaction, we consider the
${}^{1}S_0$ channel, which exhibits strong attraction
and supports a shallow bound state.
The corresponding local potential is parametrized
as a sum of three Gaussian terms, with the parameters
listed in Table~\ref{tab:Omega_potentials2},
and is given by~\cite{OmegaOmega2018}
\begin{equation}
V_{\Omega\Omega}(r)
=
\sum_{j=1}^{3}
c_j
\exp\left[-\left(\frac{r}{d_j}\right)^2\right].
\label{OmegaOmega}
\end{equation}
The parameters and their uncertainties, as determined in Refs.~\cite{Iritani2019,Zhang2026QCD2026,OmegaOmega2018}, are listed in
Table~\ref{tab:Omega_potentials2}.  In the present calculations, however, we use the central values of the potentials.

The $\Omega\Omega$ potential exhibits
short-range repulsion and intermediate-range attraction,
with the latter sufficiently strong to support a weakly
bound dibaryon. The Pauli principle restricts the allowed
$S$-wave states of two identical $\Omega$ baryons to
${}^{1}S_0$ and ${}^{5}S_2$.
The HAL QCD results~\cite{OmegaOmega2018}
indicate a strongly attractive interaction in the
${}^{1}S_0$ channel.
Unlike the nucleon--nucleon interaction, the
$\Omega\Omega$ interaction has no long-range
one-pion-exchange contribution.
In calculations of the $\Omega\Omega N$ system, this potential is combined
with the $\Omega N$ interactions described
above to examine the interplay between pairwise
interactions and few-body correlations in the
$S=-6$ sector.

\begin{table*}[!b]
\centering
\small

\caption{
Three-body $\Omega NN$ system with $T=0$, calculated with
and without the Coulomb interaction using the HAL QCD $\Omega N$
potentials of Refs.~\cite{Iritani2019,Zhang2026QCD2026}
at Euclidean time slices $t/a=11$--$16$. The $NN$ interaction is described by the Malfliet–Tjon potential~\cite{Malfliet1969,MTcorr}.
As in Tables \ref{tab:OmegaNN_T1} and \ref{tab:OmegaOmegaN_combined} results obtained with the potential of Ref.~\cite{Iritani2019}
are taken from Ref.~\cite{Lazauskas2026OmegaNNNN}.
The quantity $\rho_3^2$ is the mean-square hyperradius, defined
as the sum of the squares of the mass-scaled Jacobi
coordinates of the three-body system.
The quantities $r_i^2$ denote the mean-square distances
of the individual baryons from the center of mass
of the $\Omega np$ system, and $\langle T_3\rangle$
is the expectation value of the kinetic energy.
}
\label{tab:ONN}

\setlength{\tabcolsep}{2.5pt}

\begin{tabular}{llcccccccccccc}
\hline\hline

$t/a$ & Coulomb
& \shortstack{$E_3$ [MeV]\\\cite{Iritani2019}}
& \shortstack{$E_3$ [MeV]\\\cite{Zhang2026QCD2026}}
& \shortstack{$\rho_3^2$\\\cite{Iritani2019}}
& \shortstack{$\rho_3^2$\\\cite{Zhang2026QCD2026}}
& \shortstack{$\langle T_3\rangle$\\\cite{Iritani2019}}
& \shortstack{$\langle T_3\rangle$\\\cite{Zhang2026QCD2026}}
& \shortstack{$r_\Omega^2$\\\cite{Iritani2019}}
& \shortstack{$r_\Omega^2$\\\cite{Zhang2026QCD2026}}
& \shortstack{$r_n^2$\\\cite{Iritani2019}}
& \shortstack{$r_n^2$\\\cite{Zhang2026QCD2026}}
& \shortstack{$r_p^2$\\\cite{Iritani2019}}
& \shortstack{$r_p^2$\\\cite{Zhang2026QCD2026}} \\

\hline

\multicolumn{14}{c}{
$\Omega np$, \qquad $J^\pi=5/2^+$
} \\

\hline

11 & No
& -19.572 & --- & 8.22 & --- & 57.7 & ---
& 0.61 & --- & 1.52 & --- & 1.52 & --- \\

& Yes
& -20.882 & --- & 8.03 & --- & 58.8 & ---
& 0.59 & --- & 1.48 & --- & 1.48 & --- \\

12 & No
& -20.000 & -19.341 & 8.07 & 8.27 & 58.8 & 57.3
& 0.59 & 0.61 & 1.49 & 1.52 & 1.49 & 1.52 \\

& Yes
& -21.323 & -20.400 & 7.88 & 8.14 & 59.9 & 58.0
& 0.57 & 0.60 & 1.46 & 1.51 & 1.46 & 1.49 \\

13 & No
& -19.421 & -18.614 & 8.17 & 8.36 & 58.4 & 57.1
& 0.60 & 0.62 & 1.51 & 1.54 & 1.51 & 1.54 \\

& Yes
& -20.740 & -19.672 & 7.97 & 8.22 & 59.6 & 57.8
& 0.58 & 0.61 & 1.47 & 1.52 & 1.47 & 1.50 \\

14 & No
& -19.854 & -16.998 & 8.16 & 8.66 & 58.0 & 55.8
& 0.60 & 0.65 & 1.51 & 1.59 & 1.51 & 1.59 \\

& Yes
& -21.167 & -18.045 & 7.97 & 8.50 & 59.1 & 56.6
& 0.58 & 0.63 & 1.47 & 1.57 & 1.47 & 1.55 \\

15 & No
& --- & -16.908 & --- & 8.90 & --- & 54.0
& --- & 0.67 & --- & 1.63 & --- & 1.63 \\

& Yes
& --- & -17.942 & --- & 8.74 & --- & 54.7
& --- & 0.65 & --- & 1.61 & --- & 1.59 \\

16 & No
& --- & -14.992 & --- & 9.82 & --- & 49.3
& --- & 0.76 & --- & 1.78 & --- & 1.78 \\

& Yes
& --- & -15.990 & --- & 9.62 & --- & 50.1
& --- & 0.65 & --- & 1.61 & --- & 1.59 \\

\hline

\multicolumn{14}{c}{
$\Omega np$, \qquad $J^\pi=3/2^+$
} \\

\hline

11 & No
& -3.004 & --- & 59.1 & --- & 18.0 & ---
& 6.60 & --- & 8.00 & --- & 8.00 & --- \\

& Yes
& -3.598 & --- & 46.2 & --- & 19.6 & ---
& 4.55 & --- & 7.21 & --- & 7.21 & --- \\

12 & No
& -3.061 & -2.964 & 57.0 & 60.9 & 18.4 & 17.8
& 6.25 & 6.62 & 7.90 & 9.05 & 7.90 & 9.05 \\

& Yes
& -3.666 & -3.529 & 45.1 & 48.6 & 20.0 & 19.2
& 4.38 & 4.79 & 7.13 & 8.82 & 7.13 & 6.74 \\

13 & No
& -2.952 & -2.821 & 61.2 & 68.6 & 17.9 & 17.2
& 6.95 & 7.79 & 7.13 & 9.77 & 7.13 & 9.77 \\

& Yes
& -3.541 & -3.362 & 47.0 & 51.9 & 19.6 & 18.7
& 4.67 & 5.32 & 7.27 & 8.91 & 7.27 & 7.23 \\

14 & No
& -3.055 & -2.564 & 57.3 & 98.6 & 18.2 & 15.7
& 6.29 & 12.3 & 7.91 & 12.5 & 7.91 & 12.5 \\

& Yes
& -3.657 & -3.047 & 45.4 & 62.3 & 19.8 & 17.4
& 4.41 & 6.96 & 7.14 & 9.64 & 7.14 & 8.54 \\

15 & No
& --- & -2.607 & --- & 91.4 & --- & 15.7
& --- & 11.20 & --- & 11.9 & --- & 11.9 \\

& Yes
& --- & -3.096 & --- & 60.6 & --- & 17.3
& --- & 6.69 & --- & 9.49 & --- & 8.36 \\

16 & No
& --- & -2.450 & --- & 133. & --- & 14.3
& --- & 17.3 & --- & 15.5 & --- & 15.5 \\

& Yes
& --- & -2.882 & --- & 73.6 & --- & 15.9
& --- & 8.65 & --- & 10.6 & --- & 9.83 \\

\hline\hline

\end{tabular}
\end{table*}
\textbf{Results and discussion}. 
The present calculations are performed by solving the configuration-space Faddeev equations. We use a partial-wave decomposition of the spin-angular wave functions and the Lagrange-mesh method to discretize the radial coordinates~\cite{Baye2015}.
The theoretical framework and numerical methods have
been detailed in our previous studies~
\cite{FilKezPRC2025,FilKezPRD2025,
LazKezFilPRD2026}. In particular, the same formalism and numerical implementation were used in our recent studies 
of $\phi$-mesic nuclei~\cite{LazKezFilPRD2026,
LazKezFilPLB2026}. Here, we summarize only the aspects relevant to the systems considered.

For the $\Omega p$ interaction, Ref.\cite{Zhang2026QCD2026} recommends including finite-charge-size corrections in the Coulomb potential,  
$V_{\mathrm{Coul}}(r)=-\alpha F_C(r)/r$ (in the point-charge limit, $F_C(r) \rightarrow1$). The correction factor $F_C(r)$ 
is obtained by convoluting the Coulomb kernel with dipole 
charge distributions~\cite{Lyu2021}, using the proton and 
$\Omega^-$ charge radii of $0.841~\mathrm{fm}$~\cite{Mohr2025} 
and $0.571~\mathrm{fm}$~\cite{Can2015}, respectively. For the other potentials, we use the standard point-charge Coulomb interaction.

Table~\ref{tab:ONN} summarizes the binding energies and spatial properties of the three-body $\Omega NN$ system with $T=0$, corresponding to the $\Omega np$ particle content.  The results obtained with the earlier HAL QCD $\Omega N$ potential~\cite{Iritani2019} are compared with those obtained using the updated physical-point interaction ~\cite{Zhang2026QCD2026}. The short range nature of the  $\Omega N$ interaction, corresponding to a range comparable to that of two-pion exchange, makes orbital excitations energetically costly. Therefore only bound states with total orbital momentum $L=0^+$ can be bound. We identify states with total angular momentum and parity $J^\pi=3/2^+$ and $J^\pi=5/2^+$.
In the  $J^\pi=5/2^+$ configuration, all three particles can interact simultaneously, making this state very compact and strongly bound. The updated potential generally produces weaker binding. For $J^\pi=5/2^+$, the binding energy decreases by $\sim 3\%$ at $t/a=12$ without Coulomb interaction, while the spatial structure remains nearly unchanged.

By contrast, the $J^\pi=3/2^+$ state state is only weakly bound relative to the $\Omega N$ or deuteron threshold. This behaviour arises from the strong spin dependence of the  $\Omega N$ interaction: in the total-spin-$3/2$ configuration, only one nucleon can interact with the $\Omega$-baryon in the active
total-spin $S=2$ channel. The weak binding makes the properties of this state, particularly its spatial structure, sensitive to the details of the baryon–nucleon interaction model.  At $t/a=14$, the mean-square hyperradius increases from $57.30$ to $98.56~\mathrm{fm}^2$, accompanied by a reduction in the binding energy from $3.055$ to $2.564~\mathrm{MeV}$. This behavior indicates that the more weakly bound $J^\pi=3/2^+$ state is considerably more sensitive to the $\Omega N$ interaction.

The Coulomb interaction enhances the binding in both channels and generally reduces their spatial extension. In the calculations employing the earlier HAL QCD potential, the Coulomb interaction is treated in the point-charge approximation, whereas finite-charge-size corrections are explicitly included with the updated interaction. Consequently, differences between the Coulomb-inclusive results reflect both the modification of the strong $\Omega N$ potential and the improved treatment of the electromagnetic interaction. Nevertheless, both HAL QCD potentials predict bound states in the investigated channels, with the $J^\pi=5/2^+$ state remaining substantially more compact and deeply bound than the $J^\pi=3/2^+$ state. Although its binding energy is approximately 2.5–3.1 MeV relative to the three-body breakup threshold, this does not necessarily establish stability against breakup into a bound two-body subsystem and a spectator.

Although the updated physical-point HAL QCD $\Omega N$
interaction reduces the two-body binding energy by
approximately a factor of two \cite{Zhang2026QCD2026} compared with the earlier
near-physical-point results~\cite{Iritani2019},
the corresponding three-body $\Omega NN$ systems exhibit
a considerably smaller reduction in binding energy,
indicating a relative robustness of three-body binding
against changes in the underlying $\Omega N$ interaction. 

The dependence of the three-body energies on the Euclidean-time separation $t/a$ differs between the two HAL QCD interactions. The earlier near-physical-point potential~\cite{Iritani2019} produces decreasing variations of $E_3$, whereas the updated physical-point potential~\cite{Zhang2026QCD2026} generally yields weaker binding at higher values of $t/a$. These differences reflect the distinct Euclidean-time dependence of the extracted $\Omega N$ potentials and may be associated with residual systematic uncertainties in their lattice-QCD determination.

For the $T = 1$ $\Omega NN$ systems, the Pauli principle requires the low-energy  $nn$, $pp$, and $np$, pairs to couple to the spin-singlet $S = 0$ channel. The lowest-energy configuration in three-body system is therefore $J^\Pi=3/2^+$. In this state, only one $\Omega N$ pair interacts in the active channel, and the system is consequently weakly bound. Table ~\ref{tab:OmegaNN_T1}  presents the three-body energies and spatial properties of the  $\Omega nn$, $\Omega np$ and  $\Omega pp$ and systems with isospin $T = 1$, calculated using the earlier near-physical-point HAL QCD $\Omega N$  potential~\cite{Iritani2019} and the updated physical-point interaction~\cite{Zhang2026QCD2026}.

The updated potential generally produces weaker binding and more
spatially extended configurations. At $t/a=12$, the binding
energies decrease by approximately $8\%$, $12\%$, and $6\%$
for $\Omega nn$, $\Omega pp$, and $\Omega np$, respectively,
accompanied by increases in their mean-square hyperradii.
As $t/a$ increases, the updated interaction generally leads to weaker binding, larger spatial extensions, and lower kinetic energies.
The $\Omega pp$ and $\Omega np$ systems are more strongly
bound than $\Omega nn$, consistent with the additional attractive
Coulomb interaction between the $\Omega^-$ baryon and the
proton(s). However, the differences between the two sets of
calculations reflect changes in both the strong $\Omega N$
interaction and the Coulomb treatment, which employs the
point-charge approximation for the earlier potential and
finite-charge-size corrections for the updated interaction.
Compared with the deeply bound $T=0$, $J^\pi=5/2^+$ state,
the $T=1$ systems exhibit substantially less strongly bound and have more diffuse spatial distributions, illustrating the pronounced spin–isospin dependence of the $\Omega NN$ systems.

\begin{table*}[!htbp]
\caption{ Three-body $J=3/2^+$ states of $\Omega NN$ systems with isospin $T=1$: 
$\Omega nn$, $\Omega pp$ and $\Omega np$. Notations are the same as in Table \ref{tab:ONN}.
%The notation is the same as in Table~\ref{tab:ONN}. Three-body $\Omega NN$ systems with $T=1$ %, calculated without the Coulomb interaction for $\Omega nn$ and  $\Omega pp$
%and $\Omega np$ systems. Notations the same as in Table \ref{tab:ONN}.
%with the Coulomb interaction included for $\Omega pp$ and $\Omega np$. The calculations employ the HAL QCD $\Omega N$ potentials of Ref.~\cite{Iritani2019} at Euclidean time slices $t/a=11,12$ and Ref.~\cite{Zhang2026QCD2026} at $t/a=12$--$16$. Listed quantities are the three-body energy $E_3$, mean-square hyperradius $\rho_3^2$, kinetic-energy expectation value $\langle T_3\rangle$, and mean-square distances $r_\Omega^2$ and $r_N^2$ from the three-body center of mass.
}
\label{tab:OmegaNN_T1}

\centering
\footnotesize
\setlength{\tabcolsep}{2.2pt}
\renewcommand{\arraystretch}{1.1}

\begin{ruledtabular}
\begin{tabular}{l*{15}{c}}

& \multicolumn{5}{c}{$\Omega nn$}
& \multicolumn{5}{c}{$\Omega pp$}
& \multicolumn{5}{c}{$\Omega np$} \\
\hline
$t/a$
& \shortstack{$E_3$\\(MeV)}
& \shortstack{$\rho_3^2$\\(fm$^2$)}
& \shortstack{$\langle T_3\rangle$\\(MeV)}
& \shortstack{$r_\Omega^2$\\(fm$^2$)}
& \shortstack{$r_N^2$\\(fm$^2$)}
& \shortstack{$E_3$\\(MeV)}
& \shortstack{$\rho_3^2$\\(fm$^2$)}
& \shortstack{$\langle T_3\rangle$\\(MeV)}
& \shortstack{$r_\Omega^2$\\(fm$^2$)}
& \shortstack{$r_N^2$\\(fm$^2$)}
& \shortstack{$E_3$\\(MeV)}
& \shortstack{$\rho_3^2$\\(fm$^2$)}
& \shortstack{$\langle T_3\rangle$\\(MeV)}
& \shortstack{$r_\Omega^2$\\(fm$^2$)}
& \shortstack{$r_N^2$\\(fm$^2$)} \\

\colrule

\multicolumn{16}{c}{
HAL QCD potential of Ref.~\cite{Iritani2019}
} \\

\colrule

11
& -2.241 & 54.1 & 18.9 & 3.37 & 10.4
& -3.239 & 47.4 & 21.1 & 2.84 & 9.23
& -2.851 & 47.9 & 20.3 & 2.96 & 9.27 \\

12
& -2.350 & 52.6 & 19.5 & 3.26 & 10.2
& -3.367 & 46.3 & 21.7 & 2.76 & 9.03
& -2.971 & 46.8 & 20.9 & 2.88 & 9.06 \\

\colrule

\multicolumn{16}{c}{
HAL QCD potential of Ref.~\cite{Zhang2026QCD2026}
} \\

\colrule

12
& -2.161 & 55.2 & 18.6 & 3.41 & 10.6
& -2.968 & 50.2 & 20.0 & 3.02 & 9.77
& -2.800 & 47.5 & 19.9 & 2.99 & 9.15 \\

13
& -1.875 & 59.5 & 17.8 & 3.71 & 11.4
& -2.664 & 53.4 & 19.4 & 3.22 & 10.4
& -2.499 & 50.5 & 19.2 & 3.20 & 9.70 \\

14
& -1.309 & 72.3 & 15.8 & 4.64 & 13.8
& -2.052 & 62.1 & 17.6 & 3.80 & 12.1
& -1.898 & 58.7 & 17.4 & 3.81 & 11.2 \\

15
& -1.404 & 70.2 & 15.5 & 4.50 & 13.4
& -2.142 & 61.0 & 17.2 & 3.74 & 11.8
& -1.992 & 57.7 & 17.0 & 3.74 & 11.0 \\

16
& -1.004 & 86.1 & 12.9 & 5.68 & 16.3
& -1.677 & 71.9 & 14.6 & 4.49 & 13.9
& -1.547 & 67.9 & 14.5 & 4.52 & 12.9 \\

\end{tabular}
\end{ruledtabular}
\end{table*}

\begin{table*}[!htbp]
\caption{
Three-body $\Omega\Omega N$ systems with
$J^\pi=1/2^+$, $ L=0$, $S=1/2$ and $J^\pi=5/2^+$, $ L=0$, $S=5/2$. Notations are the same as in Table \ref{tab:ONN}. %, calculated with the Coulomb interaction included for $\Omega\Omega n$ and $\Omega\Omega p$. The calculations employ two HAL QCD $\Omega N$ potentials: the potential of Ref.~\cite{Iritani2019} at Euclidean time slices $t/a=11$--$14$, and the potential of Ref.~\cite{Zhang2026QCD2026} at $t/a=12$--$16$. 
The two-body $\Omega\Omega$ ($S=0$) ground-state energy with Coulomb interaction is $E_{\Omega\Omega}=-0.5977~\mathrm{MeV}$. %Listed quantities are the three-body energy $E_3$, kinetic-energy expectation value $\langle T_3\rangle$, mean-square hyperradius $\rho_3^2$, and mean-square distances $r_N^2$ and $r_\Omega^2$ from the three-body center of mass.
}
\label{tab:OmegaOmegaN_combined}

\centering
\footnotesize
\setlength{\tabcolsep}{3pt}
\renewcommand{\arraystretch}{1.1}

\begin{ruledtabular}
\begin{tabular}{l*{10}{c}}

& \multicolumn{5}{c}{$\Omega\Omega n$}
& \multicolumn{5}{c}{$\Omega\Omega p$} \\

\hline

$t/a$,\quad Ref.
& \shortstack{$E_3$\\(MeV)}
& \shortstack{$\langle T_3\rangle$\\(MeV)}
& \shortstack{$\rho_3^2$\\(fm$^2$)}
& \shortstack{$r_n^2$\\(fm$^2$)}
& \shortstack{$r_\Omega^2$\\(fm$^2$)}
& \shortstack{$E_3$\\(MeV)}
& \shortstack{$\langle T_3\rangle$\\(MeV)}
& \shortstack{$\rho_3^2$\\(fm$^2$)}
& \shortstack{$r_p^2$\\(fm$^2$)}
& \shortstack{$r_\Omega^2$\\(fm$^2$)} \\

\colrule

\multicolumn{11}{c}{$\Omega\Omega N,\qquad J^\pi=1/2^+,\quad L=0,\quad S=1/2$} \\

\colrule

%\multicolumn{11}{c}{HAL QCD potential of Ref.~\cite{Iritani2019}} \\

11 \quad\cite{Iritani2019}
& -4.940 & 31.8 & 19.9 & 3.78 & 1.72
& -6.925 & 35.1 & 17.0 & 3.12 & 1.50 \\

12
& -5.111 & 32.6 & 19.4 & 3.65 & 1.69
& -7.125 & 36.0 & 16.6 & 3.03 & 1.48 \\

13
& -4.834 & 32.1 & 20.0 & 3.80 & 1.73
& -6.828 & 35.5 & 17.0 & 3.12 & 1.50 \\

14
& -5.072 & 32.2 & 19.6 & 3.70 & 1.70
& -7.069 & 35.4 & 16.8 & 3.12 & 1.50 \\

\colrule

%\multicolumn{11}{c}{HAL QCD potential of Ref.~\cite{Zhang2026QCD2026}} \\

12 \quad \cite{Zhang2026QCD2026}
& -4.829 & 31.5 & 20.2 & 3.85 & 1.74
& -6.504 & 33.8 & 17.8 & 3.30 & 1.56 \\

13
& -4.464 & 30.9 & 20.9 & 4.06 & 1.79
& -6.122 & 33.4 & 18.3 & 3.43 & 1.60 \\

14
& -3.703 & 29.2 & 23.2 & 4.69 & 1.94
& -5.310 & 31.9 & 19.7 & 3.82 & 1.69 \\

15
& -3.764 & 28.4 & 23.5 & 4.75 & 1.96
& -5.353 & 30.9 & 20.0 & 3.89 & 1.72 \\

16
& -3.092 & 25.1 & 27.4 & 5.83 & 2.21
& -4.586 & 27.7 & 22.7 & 4.60 & 1.90 \\

\colrule

\multicolumn{11}{c}{$\Omega\Omega N,\qquad J^\pi=5/2^+,\quad L=0,\quad S=5/2$} \\

\colrule

%\multicolumn{11}{c}{HAL QCD potential of Ref.~\cite{Iritani2019}} \\

11 \quad\cite{Iritani2019}
& -1.848 & 18.0 & 84.3 & 8.20 & 9.26
& -3.433 & 22.3 & 49.6 & 5.21 & 5.41 \\

12
& -1.984 & 18.9 & 78.0 & 7.64 & 8.57
& -3.609 & 23.3 & 47.1 & 4.95 & 5.14 \\

13
& -1.728 & 17.9 & 89.8 & 8.65 & 9.88
& -3.307 & 22.5 & 50.9 & 5.32 & 5.56 \\

14
& -1.968 & 18.5 & 79.1 & 7.75 & 8.69
& -3.576 & 22.8 & 47.8 & 5.03 & 5.21 \\

\colrule

%\multicolumn{11}{c}{ HAL QCD potential of Ref.~\cite{Zhang2026QCD2026}} \\

12 \quad \cite{Zhang2026QCD2026}
& -1.747 & 17.5 & 87.9 & 8.71 & 9.70
& -3.088 & 20.8 & 55.0 & 5.81 & 6.03 \\

13
& -1.410 & 16.1 & 110. & 10.6 & 12.2
& -2.699 & 19.8 & 62.1 & 6.49 & 6.81 \\

14
& -0.788 & 12.5 & 225. & 19.9 & 25.1
& -1.935 & 17.0 & 85.4 & 8.72 & 9.41 \\

15
& -0.887 & 12.4 & 194. & 17.6 & 21.6
& -2.041 & 16.6 & 82.5 & 8.49 & 9.08 \\

16
& -0.507 & 8.43 & 467. & 36.0 & 52.1
& -1.483 & 12.7 & 137. & 12.6 & 14.9 \\

\end{tabular}
\end{ruledtabular}
\end{table*}

Interestingly, the three-body binding energies obtained at $t/a=15$--$16$ with the updated physical-point HAL QCD $\Omega N$ potential~\cite{Zhang2026QCD2026} closely agree with our earlier results~\cite{Lazauskas2026OmegaNNNN} based on the meson-exchange $\Omega N$ interaction of Sekihara \textit{et al.}~\cite{Sekihara2018}.

%It is interesting to mention that the three-body binding energies obtained with the updated physical-point HAL QCD potential $\Omega N$~\cite{Zhang2026QCD2026} show the closest agreement with our previous calculations~\cite{Lazauskas2026OmegaNNNN}, performed using the meson-exchange interaction $\Omega N$ of Sekihara \textit{et al.}~\cite{Sekihara2018}, at $t/a=15$--$16$.

The properties of double-$\Omega$ nuclei are similar  in some respects to those of the single-$\Omega$ systems. The strong spin dependence of the $\Omega$-baryon interactions again plays an important role in determining the properties of the bound states. As in the $\Omega NN$ systems, configurations with total orbital angular momentum $L=0^+$ are energetically favourable. Two total-angular-momentum channels support bound states: $J^\pi=1/2^+$ and $5/2^+$, see  Table~\ref{tab:OmegaOmegaN_combined}. The $1/2^+$ configuration is 
strongly bound because all three particles can interact simultaneously, taking advantage of the strongly attractive spin-singlet $\Omega\Omega$ interaction. 

The results for $\Omega\Omega n$ and $\Omega\Omega p$ systems obtained using the earlier near-physical-point HAL QCD $\Omega N$ potential~\cite{Iritani2019} and the
updated physical-point interaction~\cite{Zhang2026QCD2026} are presented in Table
~\ref{tab:OmegaOmegaN_combined}.
The updated potential generally produces weaker binding and
more spatially extended configurations. At $t/a=12$, the
binding energies decrease by approximately $5\%$ and
$9\%$ for $\Omega\Omega n$ and $\Omega\Omega p$,
respectively, in the $J^\pi=1/2^+$ channel, compared with
reductions of $\sim 12\%$ and $14\%$ in the $J^\pi=5/2^+$
channel.

The $\Omega\Omega N$ states with $J^\pi=5/2^+$ are only weakly bound because, in this configuration, the $\Omega\Omega$ pair  can not form S-wave singlet state. Their weak binding makes the properties of these spatially extended states strongly dependent on the interaction model.  The $\Omega\Omega p$ systems exhibit stronger binding than
their neutron counterparts, consistent with the additional
attractive Coulomb interaction between the proton and the
two $\Omega^-$ baryons. 
Interestingly, at $t/a=16$, the calculated energy of the
$\Omega\Omega n$ state with $J^\pi=5/2^+$ 
is $E_3=-0.507~\mathrm{MeV}$, which is less bound than the
$E_{\Omega\Omega}=-0.5977~\mathrm{MeV}$ spin-singlet state.
Within the present model, however, this state is stable because the Hamiltonian (free from tensor or spin-orbit interactions) conserves total spin and orbital angular momentum separately, thereby preventing direct decay of the three-body state into the $\Omega\Omega$ spin-singlet state. 

\textbf{Conclusions.}  
We have investigated the three-body $\Omega NN $ and $\Omega\Omega N $ systems by solving the configuration-space Faddeev equations with the updated physical-point HAL QCD $\Omega N $  interaction. Finite-charge-size corrections to the Coulomb interaction are included, and the results are compared with those obtained using the earlier near-physical-point potential~\cite{Iritani2019}. Although the updated interaction reduces the two-body $\Omega N$ binding energy by approximately a factor of two, our calculations reproduce the the earlier predictions of bound
three-body systems. The three-body binding energies are less sensitive to the change in the $\Omega N$ interaction than is the shallow two-body $\Omega N$ state, particularly in the $J^\pi=5/2^+$ channel. The more deeply bound configurations are relatively insensitive to the updated interaction, whereas the weakly bound states show a stronger dependence on the Euclidean-time separation and move closer to the corresponding breakup thresholds. In the $\Omega\Omega N$ systems, the $J^\pi=1/2^+$ states are more strongly bound and more compact than the $J^\pi=5/2^+$  ones, which are particularly sensitive to the updated interaction and to the Euclidean-time separation. These results indicate that three-body binding can remain relatively robust under substantial changes in the $\Omega N$  interaction, while weakly bound states close to breakup thresholds are much more sensitive to the details of the interaction models.

\end{document}